\documentclass[aps,prd,superscriptaddress,twocolumn, showpacs,preprintnumbers,nofootinbib, floatfix,letterpaper]{revtex4}   
\usepackage[utf8]{inputenc}
\usepackage{latexsym,amssymb,amsmath,amsfonts}
\usepackage[utf8]{inputenc}
\usepackage[usenames, dvipsnames]{color}
\usepackage{graphicx}
\graphicspath{{fig/}}
\usepackage[breaklinks, backref, colorlinks=true]{hyperref}
\usepackage{cleveref}
\usepackage{natbib}

\begin{document}
 
\title{Stochastic gravitational wave background mapmaking using regularised deconvolution}

\author{Sambit Panda}
\email{f2015467@pilani.bits-pilani.ac.in}
\affiliation{Birla Institue of Technology and Science (BITS) Pilani, Pilani, Rajasthan 333031, India}

\author{Swetha Bhagwat}
\email{swetha.bhagwat@roma1.infn.it}
\affiliation{Dipartimento di Fisica, Sapienza Universita di Roma, Piazzale Aldo Moro 5, 00185, Roma, Italy}
\affiliation{Syracuse University, Syracuse, NY 13244, USA}

\author{Jishnu Suresh}
\email{jishnu@icrr.u-tokyo.ac.jp}
\affiliation{Institute for Cosmic Ray Research (ICRR), KAGRA Observatory, The University of Tokyo, Kashiwa City, Chiba 277-8582, Japan}
\affiliation{Inter-University Centre for Astronomy and Astrophysics (IUCAA), Pune 411007, India}

\author{Sanjit Mitra}
\email{sanjit@iucaa.in}
\affiliation{Inter-University Centre for Astronomy and Astrophysics (IUCAA), Pune 411007, India}

\begin{abstract}

Obtaining a faithful source intensity distribution map of the sky from noisy data demands incorporating known information of the expected signal, especially when the signal is weak compared to the noise. We introduce a widely used procedure to incorporate these priors through a Bayesian regularisation scheme in the context of map-making of the anisotropic stochastic GW background (SGWB). Specifically, we implement the quadratic form of regularizing function with varying strength of regularization and study its effect on image restoration for different types of the injected source intensity distribution in simulated LIGO data. We find that regularization significantly enhances the quality of reconstruction, especially when the intensity of the source is weak, and dramatically improves the stability of deconvolution. We further study the quality of reconstruction as a function of regularization constant. While in principle this constant is dependent on the data set, we show that the deconvolution process is robust against the choice of the constant, as long as it is chosen from a broad range of values obtained by the method presented here.
\end{abstract}


\maketitle

\section{Introduction}

A stochastic gravitational wave background (SGWB)~\cite{1997rggr.conf..373A,1997PhRvD..56..545A} can arise from the gravitational waves (GW) produced by unresolved astrophysical and cosmological sources. For instance, a large number of unresolved distant compact binary coalescences ~\cite{2018PhRvL.120i1101A} and millisecond pulsars in galaxy clusters~\cite{2011PhRvD..84h3007D,Romano2017,HUGHES201486,2018arXiv180710620C} can produce an SGWB. The background can be significantly anisotropic due to the non-uniform distribution of astrophysical sources in the local universe~\cite{2014PhRvD..89h4076M}.
Based on the rates estimated from the observed GW signals, the detection of SGWB created by compact binaries seems promising in the near future ~\cite{2018PhRvL.120i1101A}. With the detectors becoming progressively sensitive, along with new detectors coming online and with the promise of multiple next-generation detectors, making the map of the SGWB-sky may soon become a reality.  The detection of an anisotropic SGWB will open an independent window to the universe and enable us to probe persistent GW sources which are electromagnetically dark.

Several algorithms have been proposed in the past to probe both isotropic and anisotropic SGWB~\cite{Michelson87,christ92,flan93,allen97,allen01,LazzariniRomano}.
Here we focus on the directed radiometer search~\cite{LazzariniWeiss,ballmer06,Mitra07}---a standard method to construct a sky-map of SGWB in a pixel-basis from the output of a network of ground-based interferometric detectors. A raw sky-map produced by this analysis is referred to as the `dirty map'.  A dirty map is a convolution of the true source map with the antenna response function of the detectors and contains additive noise. To obtain an estimate of the true source distribution in the sky, known as `clean map' one needs to undo the effects introduced by the detector's antenna pattern in the dirty map through a procedure called deconvolution or `cleaning'. However, the process of deconvolution, although mathematically well defined, need not be numerically stable and the procedure itself may introduce an additional source of error. In the case of SGWB mapmaking, it is known that the deconvolution process introduces a considerable amount of noise in the cleaned map because of insensitive modes of the beam. Consequently, it has been a common practice to use the dirty map directly to obtain scientific results. Here we propose a scheme to efficiently obtain clean maps using `regularised' deconvolution by incorporating prior knowledge of the signal in a Bayesian framework ~\cite{suyu,Rakhmanov:2006qm,Searle:2007uv}. 

Since the source map is unknown {\it a priori}, one needs a model independent prescription for implementing the regularisation scheme into the mapmaking procedure. Our method uses broad features of the expected characteristics of the source and the noise as priors. Incorporating the priors in this specific context is algebraically equivalent to adding a `regularisation function' to the log-Likelihood. Motivated by earlier successful attempts~\cite{suyu,Rakhmanov:2006qm,Searle:2007uv}, we use quadratic forms for the regularisation function. In particular, the regularization function we implement are constructed such that they penalize the `norm' or the `gradient' of the cleaned map. We demonstrate mapmaking using this procedure for two specific cases - a point source and an extended source closely mimicking the diffuse part of the Milky Way galaxy. In addition, one needs to make a choice on how much the clean map is allowed to depend on prior knowledge. This choice is coded in a quantity called `regularization constant', that balances the credence on the data and the prior. We also demonstrate that our procedure is not very sensitive to the exact value of regularization constant, provided it is picked from a range decided by following the prescription presented in this paper.

The paper is organized as follows. Section~\ref{sec:gwrad} briefly reviews the GW radiometer analysis. Section~\ref{sec:deconv} provides a detailed description of sky map reconstruction using regularized deconvolution. In section~\ref{sec:implement}, we discuss the numerical implementation for in our study. In Section~\ref{sec:result}, we present the results and show that the regularisation procedure improves the quality of deconvolution statistically. In section~\ref{sec:concl}, we discuss the implications of our method on SGWB mapmaking and its immediate applicability to the data from the current detectors. 
%__________________________________________________________
\section{GW RADIOMETER ANALYSIS}
\label{sec:gwrad}

The GW radiometer analysis~\cite{a_synthesis,LazzariniWeiss,ballmer06,Mitra07} is currently the standard technique to probe an anisotropic SGWB produced by unresolved astrophysical and cosmological sources. Its fundamental principle is based on the assumption that noise in geographically well-separated detectors are uncorrelated. Therefore, one can use the cross-correlation between detector outputs as a statistical measure of the stochastic GW signals. We use the cross-correlation statistic obtained by applying a direction dependent filter that accounts for the phase delay in the arrival of the GW signal at the detector locations, which is adjusted as the earth rotates. The statistic is averaged over the time segments with inverse noise weights to improve the signal-to-noise ratio (SNR)~\cite{ballmer06,Mitra07}. The filter is optimized on small time segments, typically $32-192$~sec long, and it also depends on the expected power spectral density (PSD) of the source and the detector noise. Further, it is generally assumed that each point source has the same PSD, and strengths in the two GW polarizations are equal. Then, the resultant dirty map can be thought of as a convolution of the true sky map with an effective kernel, often called as the beam function ($B$). It can be expressed as,
\begin{equation}
    D(\hat{\Omega}) = \int d(\hat{\Omega'}) B(\hat{\Omega},\hat{\Omega'}) S(\hat{\Omega'})+ n(\hat{\Omega}) \, ,
\end{equation}
where $S$ is the source map and $n$ is the noise. The beam captures the effect in the direction $\hat{\Omega}$ due to a source in the direction $\hat{\Omega'}$ and can be written as, 
\begin{widetext}
\begin{equation}
    B(\hat{\Omega},\hat{\Omega'}) = \Lambda (\hat{\Omega}) \int dt \int_{\infty} ^{- \infty} df \frac{H^{2}(f)} {{P_1(t;f) P_2(t;f)}}  \sum_{A'} F^{A'}_{1}(\hat{\Omega'},t) F^{A'}_{2}(\hat{\Omega'},t) \sum_{A} F^{A}_{1}(\hat{\Omega},t) F^{A}_{2}(\hat{\Omega},t) e^{2\pi i f \frac{{\hat \Omega}\cdot {{\Delta \mathbf{x}} (t)}}{c}} \, ,
\end{equation}
\end{widetext}
where, $\Lambda$ is a normalization constant, $H(f)$ is the source PSD, $P_{1,2}(t;f)$ are the one-sided PSD of detector noise and $\Delta \mathbf{x}(t) = \mathbf{x}_2 (t) - \mathbf{x}_1 (t)$ is the geometrical separation between the detectors located at $\mathbf{x}_1 (t)$ and $\mathbf{x}_2(t)$ respectively. The polarizations are denoted by $A = {+,\times}$ and $F^A _ {1,2} (\hat{\Omega},t)$ are the antenna functions of the detectors.

In practice, the convolution is done in a discrete basis. An extended source is modeled as a smooth collection of point sources, i.e., a collection of bright pixels. The maps are thus represented by vectors living in an $n_\text{pix}$-dimensional space, where $n_\text{pix}$ is the number of pixel in the map, and the beam becomes a matrix. Further, one can normalise the map in such a way~\cite{Mitra07} that the beam matrix becomes symmetric and the pixel-to-pixel noise covariance matrix becomes proportional to the beam matrix, which makes it convenient for numerical manipulations. The pixelised dirty map $\mathcal{D}$ can then be cast as a set of linear convolution equations,
\begin{equation}
\label{eq: convol-matrix}
\mathcal{D} = \mathcal{B} \cdot \mathcal{S} + \mathbf{n} \, ,
\end{equation}
where $\mathcal{S}$ is the pixelised source map and $\mathbf{n}$ represents (convolved) detector noise. The $i^\text{th}$ row of the matrix $\mathcal{B}$ is the response function of the radiometer when pointed to the $i^\text{th}$ pixel in the sky. Obtaining an optimal estimate of $\mathcal{S}$,  the `clean' map ($\hat{S}$), from the above convolution equation is the primary goal of this paper.

\begin{figure}[h]
    \centering
    \includegraphics[width=0.5\textwidth]{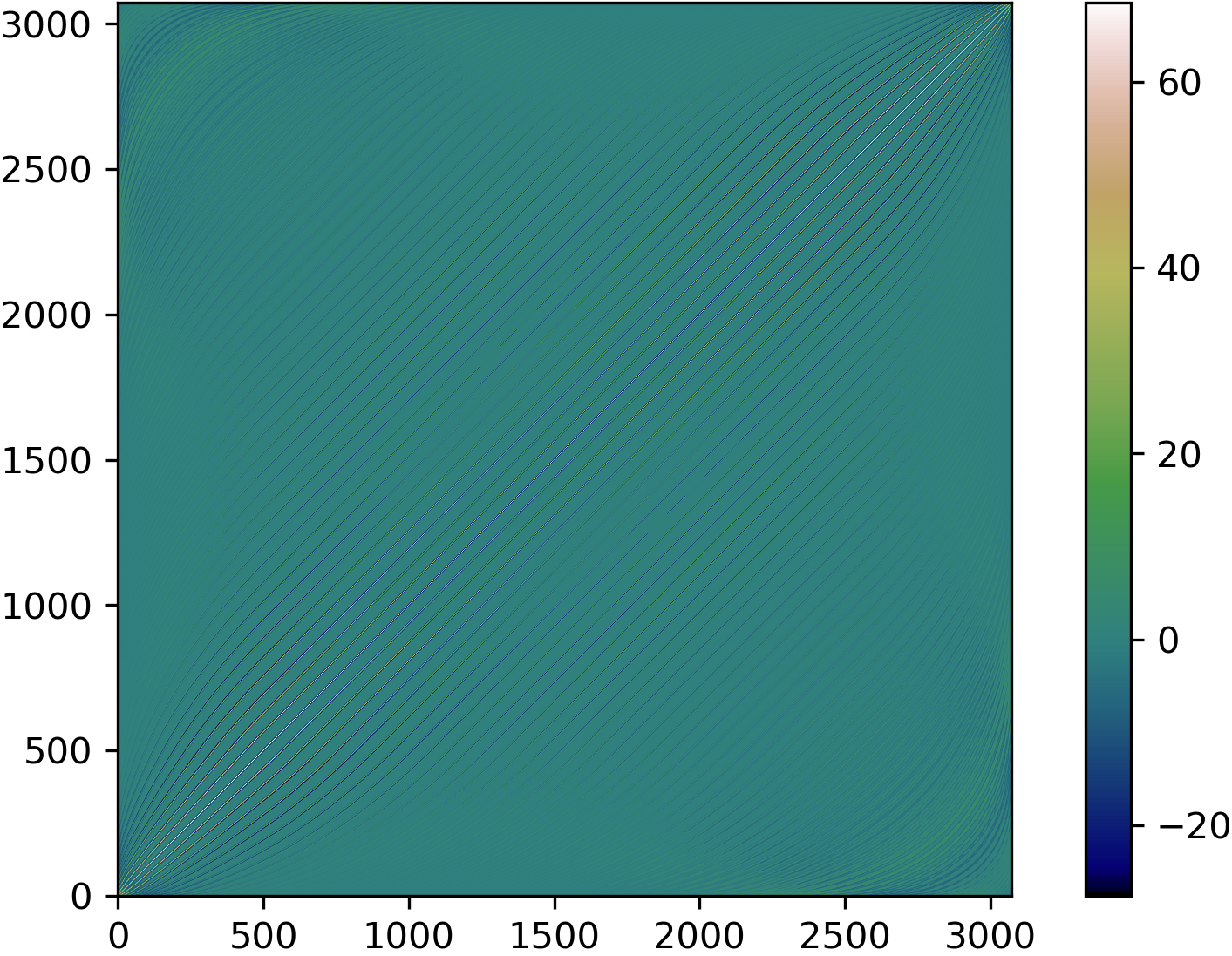}
    \caption{A characteristic beam matrix for the LIGO Hanford-Livingston baseline at $\sim 3^\circ$ pixel size (HEALPix\protect\footnote{\url{http://healpix.sf.net}} $n_{\mathrm{nside}} = 16$). Each row of the matrix is the antenna response function for the pointing direction. Since the GW radiometer receives maximum contribution from the pointing direction, this matrix is dominated by the diagonal elements. Here the stripes are related to the isoLatitude pixelization scheme - the indices of the neighboring pixels at different latitudes differ by the total number of pixels on that latitude. The beam also takes negative values.}
\label{fig:beam}
\end{figure}

A graphical representation of a typical beam matrix for the baseline created by the two LIGO detectors with the sky tessellated in $n_\text{pix} = 3072$ pixels is depicted in Fig.~\ref{fig:beam}; the colour bar indicating the values of the matrix elements. While the matrix is diagonally dominated, indicating that the maximum contribution in a map comes from the corresponding pixel in the source map, there is a significant number of off-diagonal elements which arise because the beam is broad.

The pixel-to-pixel noise covariance matrix of the dirty map has a non-trivial form. With the choice of normalisation mentioned above, the noise covariance matrix $\mathcal{N}$ becomes proportional to $\mathcal{B}$ and can be written as,
\begin{equation}
\label{eq:noise-covariance-eq}
\mathcal{N} = \kappa \, \mathcal{B} \, ,
\end{equation}
where $\kappa$ is a constant that depends on the exact normalisation convention being used, which can involve the duration of each time segment and frequency bin size.~\footnote{The qualitative features of our analysis is independent of the exact value of the proportionality constant $\kappa$. The optimum regularisation constant, which is numerically determined, absorbs this factor.}

\section{Regularized Deconvolution}
\label{sec:deconv}

To obtain an estimate for the true skymap, $\hat{S}$, the direct inversion of the convolution equation,
\begin{equation}
    \hat{S} = \mathcal{B}^{-1} \mathcal{D} \, ,
    \label{eq:direct_inversion}
\end{equation}
boosts the noise associated with the low sensitivity modes of the beam, resulting in a sky reconstruction dominated by numerical noise. This can be understood by looking at the very low singular values of the matrix $\mathcal{B}$ by performing a singular value decomposition (SVD)~\cite{Mitra07}. This leads to a small condition number (that is, the beam is ill-conditioned), rendering the direct inversion unreliable. Therefore, given a $\mathcal{D}$, there may not be a unique numerical solution $\hat{S}$. The weak intensity of the SGWB signal with respect to the detector noise level further enhances this issue. As a result, a direct inversion of the convolution equation does not lead to a reliable solution.

However, problems of this nature exist in many fields and a traditional way to tackle this has been through incorporating prior information about the system. For instance, a similar discussion on image reconstruction could be found in the context of strong gravitational lensing in \citet{suyu}. In a Bayesian framework, one can implement various regularization schemes as priors and incorporating them in the deconvolution routine is expected to remedy the ill-posed nature of Eq.~(\ref{eq:direct_inversion}).

\subsection{Most Likely Solution : $\mathcal{S}_{ml}$}

In a Bayesian framework, the problem can be formulated as finding the most likely solution. Assuming a Gaussian noise model the likelihood function is given by,
\begin{equation}
\label{eq:likelihood}
P(\mathcal{D}|\mathcal{S},\mathcal{B}) = \frac{1}{\mathcal{Z_L}} e^{{-\frac{1}{2}}(\mathcal{D} -\mathcal{B} \mathcal{S})^T N^{-1} (\mathcal{D} -\mathcal{B}\mathcal{S})}~.
\end{equation}
Here $\mathcal{Z_L}$ is the normalization constant. The signal that maximizes the above likelihood function gives an estimate $\hat{S}$, which is called the most likely solution $\mathcal{S}_{ml}$. 

Another perspective to this problem can be obtained by looking at the $\chi^2$. Using the standard $\chi^2$ estimator, the likelihood function can be written as,
\begin{equation}
P(\mathcal{D}|\mathcal{S},\mathcal{B}) = \frac{1}{\mathcal{Z_L}} e^{- \chi^2} \, ,
\end{equation}
and $\mathcal{S}_{ml}$ can be interpreted as  the solution that minimizes the $\chi^2$,
\begin{equation}
\nabla \chi ^2(\mathcal{S}_{ml}) = 0 \, . \label{eq:chisqmin}
\end{equation}
Solving the above equation one gets,
\begin{equation}
\label{eq:sml}
\mathcal{S}_{ml} =( \mathcal{B}^T N^{-1} \mathcal{B})^{-1} ( \mathcal{B}^T N^{-1}\mathcal{D}) \,.
\end{equation}
Substitution of $\mathcal{N} = \kappa \mathcal{B}$ then reduces the above formula to Eq.~(\ref{eq:direct_inversion}). Thus, likelihood maximization in the context of SGWB map-making reduces to an ill-posed problem.
This is because the set of linear equations that describe deconvolution are not independent, i.e., the number of free parameters in the problem is much more than the number of constraint equations. This invokes the risk of over-fitting by minimizing $\chi^2$ to an unrealistically small value. One has to incorporate regularization for a reliable minimization of $\chi^{2}$. Regularization serves as a set of additional constraints to the set of linear equations, Eq.~(\ref{eq: convol-matrix}), and prevents over-fitting of data.

\subsection{Most Probable Solution : $\mathcal{S}_{mp}$}

In a Bayesian framework, the prior distribution function incorporates additional information on the signal in the form of a function $R(\mathcal{S})$ which we refer to as the regularization function. Given $R(\mathcal{S})$ and a regularization strength $\mu$, the prior distribution can be written as,
\begin{equation}
P(\mathcal{S}|R(\mathcal{S}),\mu) = \frac{1}{\mathcal{Z_P}}e^{-\mu R(\mathcal{S})} \, ,
\end{equation}
where $\mathcal{Z_P}$ is the normalization for the probability. Using Bayes theorem, the posterior probability of getting the signal $\mathcal{S}$, given the dirty map $\mathcal{D}$, the beam $\mathcal{B}$ and the regularization function $R(\mathcal{S})$ is as follows,
\begin{equation}
P (\mathcal{S}|\mathcal{D}, \mathcal{B}, \mu, R(\mathcal{S})) = \frac{P (\mathcal{D}|\mathcal{S},\mathcal{B})~P(\mathcal{S}|R(\mathcal{S}), \mu)}{P(\mathcal{D}|\mu,\mathcal{B},R(\mathcal{S}))} \, .
\end{equation}
Unlike $P(\mathcal{D}|\mathcal{S},\mathcal{B})$, which depends only on data $\mathcal{D}$, $P (\mathcal{S}|\mathcal{D}, \mathcal{B}, \mu, R(\mathcal{S}))$ evaluates the probability of solution $\mathcal{S}$ by combining the information in both $\mathcal{D}$ and $R(\mathcal{S})$, through the posterior distribution,
\begin{equation}
P (\mathcal{S}|\mathcal{D}, \mathcal{B}, \mu, R(\mathcal{S})) = \frac{1}{\mathcal{Z_P}}e^{-M(\mathcal{S})} \,.
\end{equation}
where $\mathcal{Z_P}$ is the Bayesian evidence. $M(\mathcal{S})$ is defined as,
\begin{equation}
M(\mathcal{S}):= \chi^2 (\mathcal{S}) +\mu R(\mathcal{S}) \, ,
\end{equation}
which contains two competing function $\chi^2$ and $R(\mathcal{S})$. $\mathcal{S}_{mp}$ is obtained by maximizing the posterior probability, that minimizes $M(\mathcal{S})$, the linear combination of these competing functions, via solving,
\begin{equation}
\nabla M(\mathcal{S}_{mp}) = \nabla (\chi^2 (\mathcal{S}_{mp}) + \mu R(\mathcal{S}_{mp})) = 0 \, . \label{eq:Mmin}
\end{equation}
This is analogous to Eq.~(\ref{eq:sml}).
Let $\lambda := \mu (\kappa \Delta t/4)$ and $\mathcal{C}$ be the Hessian of the regularization function $R(\mathcal{S})$, then it can be shown that $\mathcal{S}_{mp}$ acquires the following simple form [c.~f. Appendix~\ref{app:smp} for the detailed derivation],

\begin{align}
    \mathcal{S}_{mp} = (\mathcal{B} + \lambda \mathcal{C})^{-1} \mathcal{D} \, . \label{eq:smp}
\end{align}
From the above equations, regularization can also be viewed as modifying the convolution kernel such that the problem becomes less ill-posed. In short, the reconstruction of the source map without a prior corresponds to evaluating the expression for $\mathcal{S}_{ml}$, while regularized deconvolution corresponds to finding $\mathcal{S}_{mp}$.

\subsection{Regularization function}
\label{sec:Regularization function}
We choose to investigate the effect of norm and gradient regularization functions on reconstruction of the source map.

\subsubsection{Norm Regularization} 
Norm regularization introduces a preferential bias towards solutions that minimizes the norm of the map. This is seen to have a noise suppression effect, especially in the case of point-like sources.
\begin{equation}
R_\text{norm} (\mathcal{S}) = \frac{1}{2} \sum_{i=1}^{n_\text{pix}} \mathcal{S}_i ^2 \, .
\end{equation}
It is clear that the Hessian matrix for this regularization form is the identity matrix,
\begin{equation}
\mathcal{C}_{ij} \ := \ \nabla_{\mathcal{S}_i} \nabla_{\mathcal{S}_j} R_\text{norm} (\mathcal{S}) \ = \ \delta_{ij} \, .
\end{equation}

\subsubsection{Gradient Regularization}
Gradient regularization incorporates a preference towards smooth source reconstruction by penalizing the intensity difference between the neighboring pixels. This is motivated by the idea that the spatial fluctuations in the intensity of the noise will be more than the spatial fluctuation in the signal, especially if the signal has an extended pattern in the sky, that is, it prefers {\em minimum variation} of intensities in the reconstructed map. The gradient is minimized by using the following form of regularization,
\begin{equation}
R_\text{grad} (\mathcal{S}) = \frac{1}{2} \sum_{i=1} ^{n_\text{pix}} \sum_{k = 1}^{n_i} (\mathcal{S}_i - \mathcal{S}_{j_{ik}}) ^2 \, ,
\end{equation}
where $j_{ik}$ is the pixel number of the $k^\text{th}$ nearest neighbour to the $i^\text{th}$ pixel and $n_i$ is the number of pixels touching the $i^\text{th}$ pixel, $n_i = 8$ if each pixel has four sides.
The Hessian for this form becomes,
\begin{equation}
\mathcal{C}_{ij} \ := \ \nabla_{\mathcal{S}_i} \nabla_{\mathcal{S}_j} R_\text{grad} (\mathcal{S}) \ = \ \delta_{ij} - 2 n_i \delta_{ij} \,.
\end{equation}
Notice that these two regularization functions do not contain any information about a specific source intensity distribution and it is in this sense that they are uninformative priors. These regularization schemes use knowledge of statistical properties of both the noise and the expected source distribution. However, we emphasize that regularization introduces a non-zero bias. Derivation of the bias estimate is presented in Appendix \ref{app:bias}. 

\subsection{The Strength of Regularization}
The strength of regularization $\lambda$ decides the relative weights between the goodness-of-fit and the bias towards our prior knowledge. It is crucial to choose a $\lambda$ that strikes an optimal balance between the two. While very low values of $\lambda$ increase the risk of over-fitting the data, setting $\lambda$ to large numbers will lead to a highly biased solution. In principle, one could systematically compute the optimal value of $\lambda$ given a data set by maximizing $\log P (\lambda|\mathcal{D}, \mathcal{B}, R(\mathcal{S}))$ with respect to $\log{\lambda}$.

In this section, we present a systematic prescription to pick an optimal value of $\lambda$, which in principle depends on the data itself.

%From Bayes' theorem,
%%\begin{equation}
%P (\lambda|\mathcal{D}, \mathcal{B}, R(\mathcal{S})) = \frac{P (\mathcal{D}|\lambda, \mathcal{B}, R(\mathcal{S}))P %%(\lambda)}{P(\mathcal{D}|R(\mathcal{S}),\mathcal{B})} \, ,
%\end{equation}
%
%%invoking uniform prior in $\log{\lambda}$, since the denominator of the above expression is independent of $\lambda$, we get,
%
%\begin{equation}
%P (\lambda|\mathcal{D}, \mathcal{B}, R(\mathcal{S})) \propto P (\mathcal{D}|\lambda, \mathcal{B}, R(\mathcal{S}))  P(\lambda) \,.
%\end{equation}
%
%Finding the optimal value of $\lambda$ then boils down to maximizing $\log P (\lambda|\mathcal{D}, \mathcal{B}, R(\mathcal{S}))$ with respect to $\log{\lambda}$,
%
%\begin{equation}
%\frac{d}{d\log{\lambda}}(\log P (\mathcal{D}|\lambda, \mathcal{B}, R(\mathcal{S}))) \ = \ 0 \, .
%\end{equation}
%
%For $\mathcal{S}_{\mathrm{reg}} = 0$, this gives a non-linear equation whose solution corresponds to the optimal value of $\lambda$. To obtain this one has to numerically solve,
%
%\begin{equation}
%\label{eq:optimal-lambda}
%2 \lambda \, R(\mathcal{S}_{mp}) = n_\text{pix} - \lambda \, \mbox{Tr}[A]^{-1} \,.
%\end{equation}

However, generally this happends to be a transcendental equation that needs to be solved iteratively. Furthermore, it involves calculating $\mathcal{S}_{mp}$ and is, therefore, computationally very expensive. Instead, for our study, we perform the deconvolution using a range of $\lambda$ and pick the $\lambda$ that maximizes the estimator of the quality of reconstruction. [c.f  the variation of the estimator as a function of $\lambda$ as presented in Fig.~\ref{fig:nsp_gradient} and \ref{fig:nmse_norm}]. However, we confirm that the $\lambda$ that maximizes the estimator of quality of reconstruction also satisfies Eq.~(\ref{eq:optimal-lambda}) up to numerical tolerance. Furthermore, in Section~\ref{sec:result} we argue that our deconvolution procedure is robust to the choice of $\lambda$ as long as it is within a range which can be determined by following the above procedure.

\subsection{Measure of quality of recovery}
There is no unique prescription to construct a quantitative estimator that measures the quality of a reconstructed map. If one considers the source map and the clean map as two vectors living in a vector space of dimension $n_\text{pix}$, a measure that quantifies the quality of the reconstructed map would be to calculate the separation between the two vectors. However, while measuring this distance, there is no unique prescription for choosing the metric on the vector space which contains these vectors. If we want to use distance as an estimator for detection of a pattern in the sky, then a good choice of metric would be one that minimizes false dismissals and false alarms. Construction of this estimator is beyond the scope of this study and will be investigated in a future work.

For our study, we construct an estimator called the Normalised Scalar Product, NSP, which quantifies the deviation of the source map and the recovered map through an inverse norm weighted Euclidean inner product. We define NSP as, 
\begin{equation}
    \text{NSP} \ = \ \frac{\mathbf{A} \cdot \mathbf{B}}{\sqrt{\|\mathbf{A}\| \|\mathbf{B}\|}} \,,
\end{equation}
where, $\mathbf{A}$, $\mathbf{B}$ in our case are the source map and the reconstructed map respectively.  
We show that this quantifies the goodness of recovery sufficiently well, in the context of this study for an extended source skymap. However, we caution the reader that it may be possible to construct better measures by a more careful choice of the metric in defining the inner product.

Furthermore, we also use the Normalised Mean Squared Error (NMSE) as another independent measure to quantify the deviation of the reconstructed map from the source map. NMSE is mathematically defined as, 
\begin{equation}
\text{NMSE} \ = \ \frac{\|\mathbf{A}-\mathbf{B}\|^2}{\|\mathbf{A}\|^2} \,.
\end{equation}

We find that NSP is a better quantifier for an extended source and NMSE is more suited for a point source. This is because NSP gives zero weight to pixels where no injections were made; hence noise in those pixels in the recovered maps are ignored, which may not be appropriate when a localised source is under consideration. On the other hand, when we are interested in an extended pattern on the sky, matching of the recovered pattern is more important than the overall normalisation, which is captured well by NSP. Needless to mention, better recovery is indicated by higher NSP or lower NMSE.
%%% TABLE %%%
\begin{table*}
\centering
%\begin{tabular}{|l|c|c|c|c|c|c|c|c|c|c|}
\begin{tabular}{lcccccccccc}
\hline\hline
Source & $\text{iter}_\text{No-reg}$ & $\lambda$ & {$\text{NSP}_\text{dirty}$} & {$\text{NSP}_\text{No-reg}$} & {$\text{NSP}_\text{Norm}$} & {$\text{NSP}_\text{Grad}$} & {$\text{NMSE}_\text{dirty}$} & {$\text{NMSE}_\text{No-reg}$} & {$\text{NMSE}_\text{Norm}$} & {$\text{NMSE}_\text{Grad}$}\\\hline
Strong extended & 12 & 5 &  0.9926 &  0.8437 &  0.8360 &  0.8798 &       0.0155 &    0.2887 &    0.3013 &    0.2259 \\\hline
Weak extended & 5 & 50 &  0.9033 &  0.6587 &  0.6448 &  0.8407 &       0.2486 &    0.6169 &    0.5851 &    0.2932 \\\hline
Very weak extended & 3 & 500 &  0.6678 &  0.3817 &  0.3533 &  0.7906 &       1.5539 &    1.3300 &    0.8752 &    0.3756 \\\hline
Strong point & 3 & 100 &  0.9384 &  0.2530 &  0.2544 &  0.2261 &    0.1447 &    0.9475 &    0.9409 &    0.9491 \\\hline
Weak point & 3 & 1000 &  0.5186 &  0.0711 &  0.0965 &  0.0820 &    3.6186 &    2.9303 &    1.0379 &    1.0249 \\\hline
Very weak point & 3 & 1000 &  0.3243 &  0.0381 &  0.0536 &  0.0547 &   14.4745 &    9.4343 &    1.2914 &    1.1830 \\\hline
\hline
\end{tabular}
\caption{Quantitative measures of goodness of reconstruction in terms of NSP and NMSE of sky-maps shown in Figures~\ref{fig:extended} and \ref{fig:point}. NSP (better measure for extended sources) and NMSE (better measure for point sources) are quoted for recovered maps obtain by no deconvolution (comparing dirty map to beam convolved injected map), unregularised deconvolution, and norm \& gradient regularised deconvolution. The number of iterations for unregularised deconvolution ($\text{iter}_\text{No-reg}$) and regularisation strength ($\lambda$) are also listed. Except for strong sources, incorporating regularization significantly improves the quality of reconstruction.}
\label{tab:nspnmse}
\end{table*}

\section{Numerical Implementation}
\label{sec:implement}

In our study, calculation of $\mathcal{B}$ is the most computationally expensive step; it is equivalent of making one dirty map for each pixel by placing a unit point source at that pixel. For this we use the software pipeline called \texttt{PyStoch} ~\cite{Ain:2018zvo} that utilizes advantages of folded data~\cite{Ain_Folding,eric_ain} and the HEALPix~\cite{HEALPix} pixelization scheme. By considering the optimal resolution required for the radiometer analysis for the two LIGO detectors at Hanford and Livingston, we choose $n_{\mathrm{side}} = 16$ in the HEALPix scheme. This corresponds to a pixel width of $\sim 3^{\circ}$ and the whole sky is tessellated into $n_\text{pix}=3072$ pixels. 

Each injection is done such that every pixel of the map is assigned a coefficient which corresponds to the intensity of that pixel in an arbitrary unit. Although arbitrary, this unit is consistent across all the plots presented in this paper. Therefore, it can be used to compare the relative strengths of the injected signals. Note that defining an SNR for these maps turns out to be tricky due to the non-trivial pixel-to-pixel covariance of the noise.

The time duration of each segment, $\Delta T$, is taken as 52~sec. The upper cut-off frequency is set to $512$~Hz. The source is assumed to have a flat PSD, $H(f) = 1$.  We follow recipes given in \citet{Ain:2018zvo} and \citet{Mitra07} to generate dirty maps with noise from injected sources.

We implement Bayesian regularized deconvolution to reconstruct the source map from the dirty map. We vary the parameter $\lambda$ over a range of $1$ to $10^6$ in logarithmic intervals and pick a value that (nearly) maximizes NSP or minimizes NMSE, as shown in Fig.~\ref{fig:nsp_gradient} and \ref{fig:nmse_norm} for different injection strengths. We later demonstrate that the qualitative results are weakly sensitive to the choice of $\lambda$. To perform the inversion of Eqs.~(\ref{eq:sml}) and (\ref{eq:smp}) we use an  in-built conjugate gradient solver ({\texttt cgs}) in the Python \texttt{SciPy} package. We set a tolerance of $10^{-6}$ for our study and a maximum iteration of $50-100$. Furthermore, to plot the sky maps, we use the Mollweid projection scheme.

\section{Results}
\label{sec:result}

We demonstrate the capabilities of our method for the extended sources as well as for the localized point-like sources with varying intensities. We apply both the gradient and the norm regularized deconvolution schemes for each of these cases to obtain the clean maps. Here we emphasize that gradient regularization is better tailored for an extended source distribution and the norm regularization scheme performs better for a point-like source. Furthermore, NSP is a better quantifier for the quality of deconvolution for an extended source, while NMSE is better suited for a point-like source. However, irrespective of the true source distribution, we present the values of both NSP and NMSE for quantifying the performance of the gradient and norm regularisation in Table~\ref{tab:nspnmse}. For brevity, we include only selected plots in the paper that correspond to (1)~Gradient regularisation for extended sources with the NSP as the quality measure and (2)~Norm regularisation for point sources with the NMSE as the quality measure. We then perform a simulation to show how the choice of $\lambda$ without any fine-tuning generically improves the quality of deconvolution.

\subsection{Extended source: Gradient regularisation}
\label{sec:extended}

\begin{figure*}
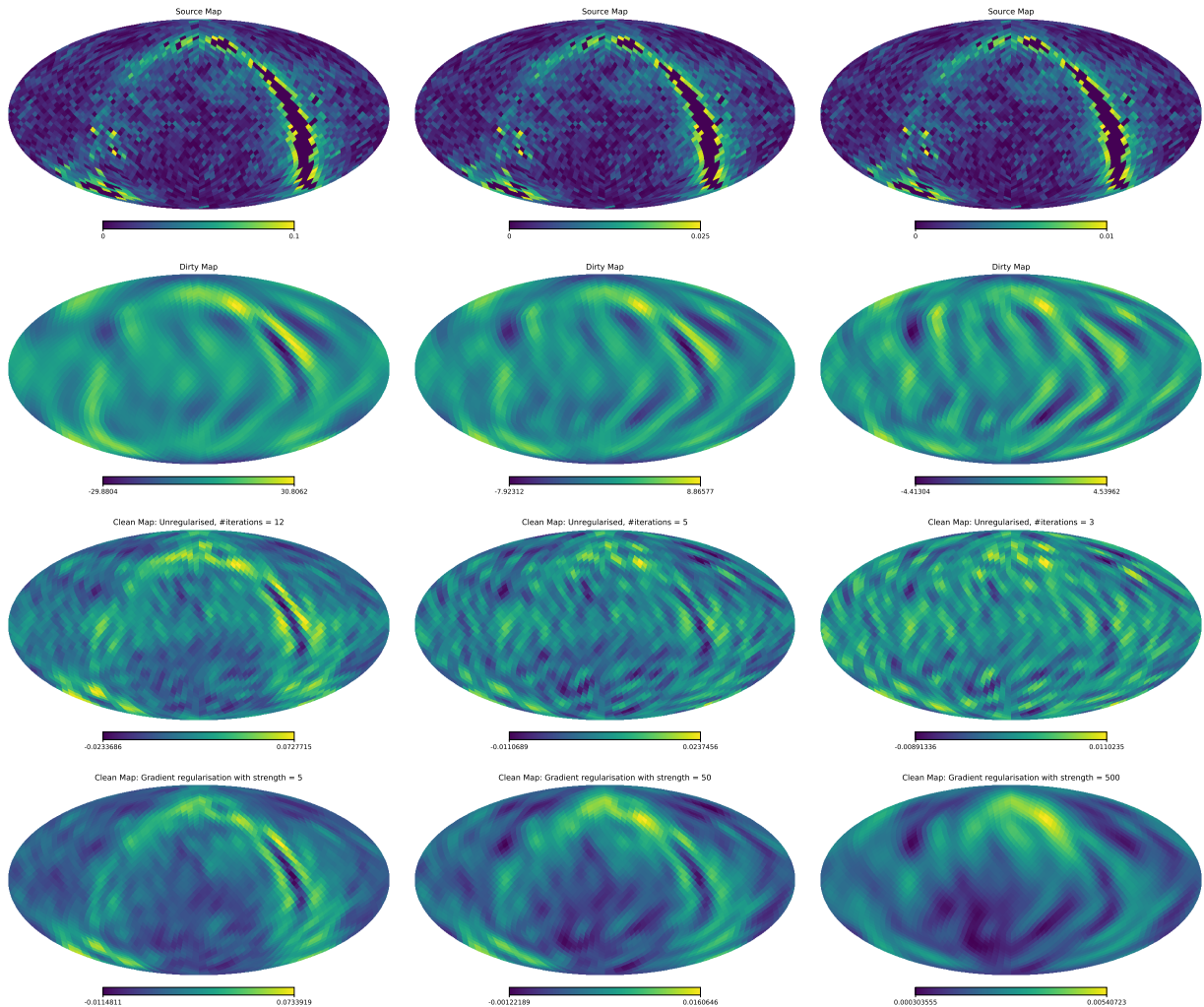

\centering
\includegraphics[width=0.3\textwidth]{{{source-ext0.1}}}
\includegraphics[width=0.3\textwidth]{{{source-ext0.025}}}
\includegraphics[width=0.3\textwidth]{{{source-ext0.01}}}\\
\includegraphics[width=0.3\textwidth]{{{dirty-ext0.1}}}
\includegraphics[width=0.3\textwidth]{{{dirty-ext0.025}}}
\includegraphics[width=0.3\textwidth]{{{dirty-ext0.01}}}\\
\includegraphics[width=0.3\textwidth]{{{clean_no-reg_ext0.1_iters12}}}
\includegraphics[width=0.3\textwidth]{{{clean_no-reg_ext0.025_iters5}}}
\includegraphics[width=0.3\textwidth]{{{clean_no-reg_ext0.01_iters3}}}\\
\includegraphics[width=0.3\textwidth]{{{clean_grad-reg_ext0.1_lambda5}}}
\includegraphics[width=0.3\textwidth]{{{clean_grad-reg_ext0.025_lambda50}}}
\includegraphics[width=0.3\textwidth]{{{clean_grad-reg_ext0.01_lambda500}}}
\caption{Illustration of gradient regularised deconvolution for extended sources. Each column corresponds to a progressively decreasing strength of injected signal with a skymap that resembles the Milky Way galaxy. The rows correspond to injected map, dirty map, unregularised and regularised clean maps. The number of iterations for the unregularised clean maps are $12, 5, 3$ respectively. For regularised deconvolution $\lambda$ was chosen as $5, 50, 500$ respectively with $100$ iterations. For a strong source, one can see that deconvolution without regularization produces good reconstruction of the true sky, regularization is less effective here. However, when the source is weak, regularisation brings dramatic improvement in deconvolution. For weak sources, the unregularised clean maps (even the dirty maps) are dominated by noise, while regularised deconvolution reduces noise and brings some of the source map features above the noise level. Quantitative measures are presented in Table~\ref{tab:nspnmse}.}
\label{fig:extended}
\end{figure*}

Here we present the results of implementing gradient regularized deconvolution for extended sources. Since the pixel-to-pixel variation of noise is expected to be much higher than that of the source, one is motivated to choose a gradient regularization scheme for recovering extended source patterns (a preference to smoother reconstructed map). The results are qualitatively presented in Fig.~\ref{fig:extended} and quantitatively in Table~\ref{tab:nspnmse}. 
\begin{figure*}
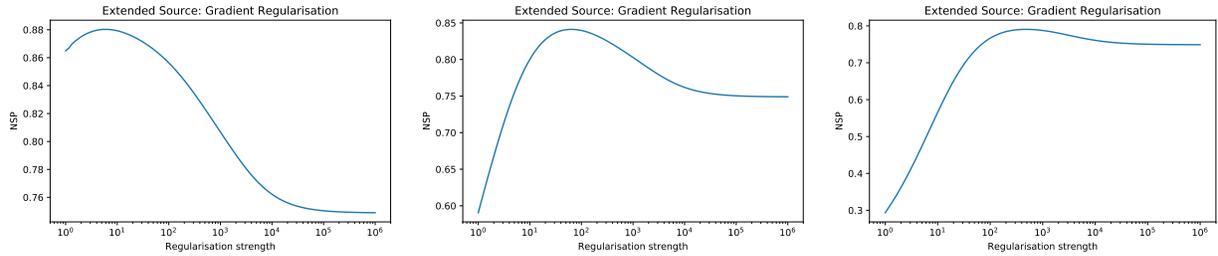

\centering
\includegraphics[width=0.3\textwidth]{{{nsp_grad_vs_lambda_ext0.1}}}
\includegraphics[width=0.3\textwidth]{{{nsp_grad_vs_lambda_ext0.025}}}
\includegraphics[width=0.3\textwidth]{{{nsp_grad_vs_lambda_ext0.01}}}
\caption{NSP (of clean map generated using gradient regularization) versus $\lambda$ for extended source with strong, weak and \textit{very} weak signal strengths respectively. The plots show that, as expected, optimal strength of regularisation, which maximises NSP, reduces with the strength of the source one is probing. However, the curves are reasonably `flat', indicating that a broad range of values of $\lambda$ can yield near-optimal result.}
\label{fig:nsp_gradient}
\end{figure*}

Each column of Fig.~\ref{fig:extended} corresponds to an increasingly diminishing injected signal strength, which can be seen in the magnitude in the colour bar of the top row of the figure indicating the pixel intensities in an arbitrary unit. The second row shows the corresponding dirty maps, and the third row shows the unregularised clean maps. Unregularised deconvolution here tends to diverge if the number of iterations is increased beyond a certain value (the value depends on the injection). We plot the NSP of the recovered map against the number of iterations and choose the number that corresponds to the best quality, i.e., maximum NSP. An example plot for a strong injection is provided in the top left panel of Fig.~\ref{fig:iter}. When the signal is very weak, unregularised deconvolution starts diverging from the first iteration, making the reconstruction unreliable. Nevertheless, we set the number of iterations to $3$ in such cases for plotting and comparison purposes.

The fourth row of Fig.~\ref{fig:extended} shows the results of regularised deconvolution. We determine the optimal regularisation constant $\lambda$ from Fig.~\ref{fig:nsp_gradient} based on the process described in Section~\ref{sec:implement}. Weaker the source, stronger the regularization necessary. This is consistent with what one expects naively. We set $\lambda = 5, 50, 500$ respectively for the injection described here. For weak sources, regularisation introduces significant improvement. The dirty and unregularised clean maps for weak injections are dominated by noise, while regularisation brings out some features of the injected source. Table~\ref{tab:nspnmse} shows this quantitatively. For instance, we see that for a very weak injection NSP increases from $0.38$ to $0.79$, while for a strong source, regularisation performs sub-optimally (sometimes worsens the result) and introduces a bias. 

\subsection{Point source: Norm regularisation}
\label{sec:point}

\begin{figure*}
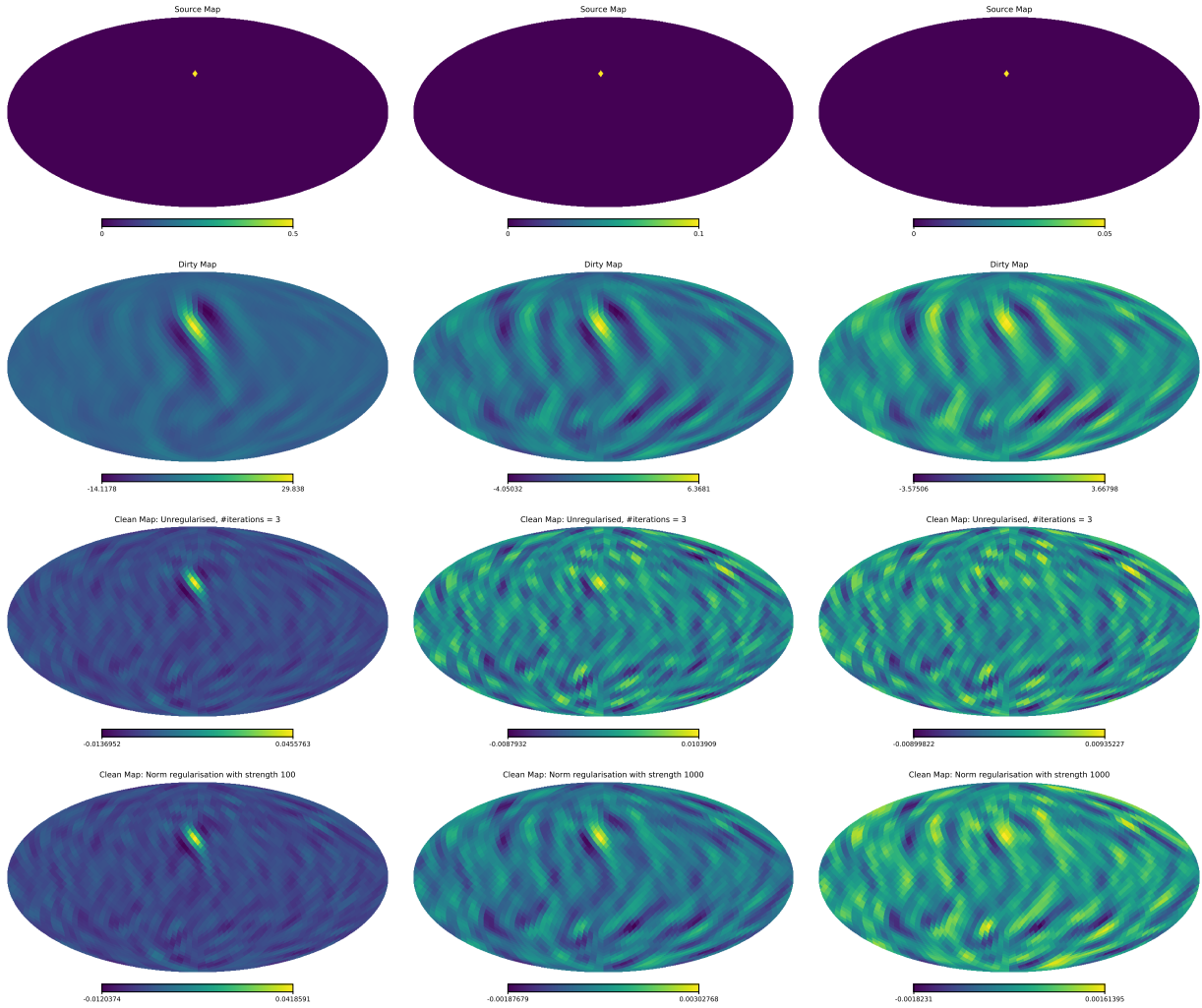

\centering
\includegraphics[width=0.3\textwidth]{{{source-point0.5}}}
\includegraphics[width=0.3\textwidth]{{{source-point0.1}}}
\includegraphics[width=0.3\textwidth]{{{source-point0.05}}}\\
\includegraphics[width=0.3\textwidth]{{{dirty-point0.5}}}
\includegraphics[width=0.3\textwidth]{{{dirty-point0.1}}}
\includegraphics[width=0.3\textwidth]{{{dirty-point0.05}}}\\
\includegraphics[width=0.3\textwidth]{{{clean_no-reg_point0.5_iters3}}}
\includegraphics[width=0.3\textwidth]{{{clean_no-reg_point0.1_iters3}}}
\includegraphics[width=0.3\textwidth]{{{clean_no-reg_point0.05_iters3}}}\\
\includegraphics[width=0.3\textwidth]{{{clean_norm-reg_point0.5_lambda100}}}
\includegraphics[width=0.3\textwidth]{{{clean_norm-reg_point0.1_lambda1000}}}
\includegraphics[width=0.3\textwidth]{{{clean_norm-reg_point0.05_lambda1000}}}
\caption{Illustration of norm regularised deconvolution for a localised source. Each column corresponds to a progressively decreasing strength of injected sky-map with a point source at an arbitrary pixel. The rows correspond respectively to injected map, dirty map and clean map without and with regularisation. The number of iterations for unregularised and regularised deconvolution are chosen as $3$ and $50$ respectively. For regularised deconvolution $\lambda$ was chosen as $100, 1000, 1000$ respectively. For strong injections, deconvolution without regularization produces a better reconstruction of the true sky-map than when regularization is implemented. For weak injections, regularisation brings dramatic improvement in deconvolution, making the source stand out in noise. Quantitative measures of these maps are presented in Table~\ref{tab:nspnmse}.}
\label{fig:point}
\end{figure*}

Norm regularization is motivated by the idea that a large part of the sky generally does not contain any source. The gradient regularization scheme is not optimal for such localized sources, as it smears out sharp features in the reconstructed maps. Here we present the results obtained from norm regularized deconvolution applied to point source injections of different intensities. The maps are presented in Fig.~\ref{fig:point} and the quantitative results are provided in Table~\ref{tab:nspnmse}. 
\begin{figure*}
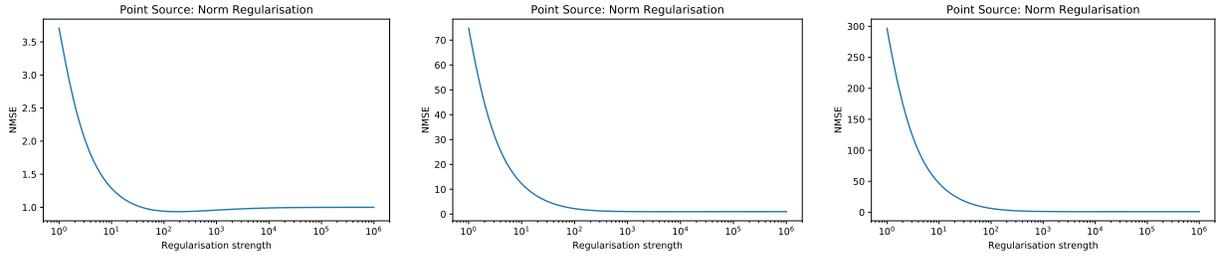

\centering
\includegraphics[width=0.3\textwidth]{{{nmse_norm_vs_lambda_point0.5}}}
\includegraphics[width=0.3\textwidth]{{{nmse_norm_vs_lambda_point0.1}}}
\includegraphics[width=0.3\textwidth]{{{nmse_norm_vs_lambda_point0.05}}}
\caption{NMSE (of clean map generated using norm regularization) versus $\lambda$ for point sources with strong, weak and \textit{very} weak signal strengths respectively. The plots show that the optimal strength of regularisation that minimises NMSE, reduces with the strength of the source. Notice that the curves are nearly `flat' beyond a certain $\lambda$, indicating that a broad ranges of values of $\lambda$ can yield near-optimal results.}
\label{fig:nmse_norm}
\end{figure*}

The first row of Fig.~\ref{fig:point} shows the injected point sources for three cases with diminishing strengths (from left to right). The second row shows the corresponding dirty maps. The third row shows clean maps obtained by unregularised deconvolution. Here the implementation of conjugate gradient is unstable and the results get saturated by numerical noise for even a small number of iterations. We therefore set the number of iteration to $3$ for all the cases here.

For regularised deconvolution, we plot NMSE versus $\lambda$ in Fig.~\ref{fig:nmse_norm}. The NMSE curve remains nearly constant for a large range of $\lambda$; this indicates that any choice of $\lambda$ above a certain critical value would produce nearly optimal results. We set $\lambda = 100$ for the strong sources and $1000$ for the weaker sources. Regularisation improves the quality of deconvolution for weak sources, as seen in the last row of Fig.~\ref{fig:point}. NMSE for a very weak point source improves by a huge factor, from $9.4343$ to $1.2914$ [c.f. Table~\ref{tab:nspnmse}]. 

\begin{figure*}
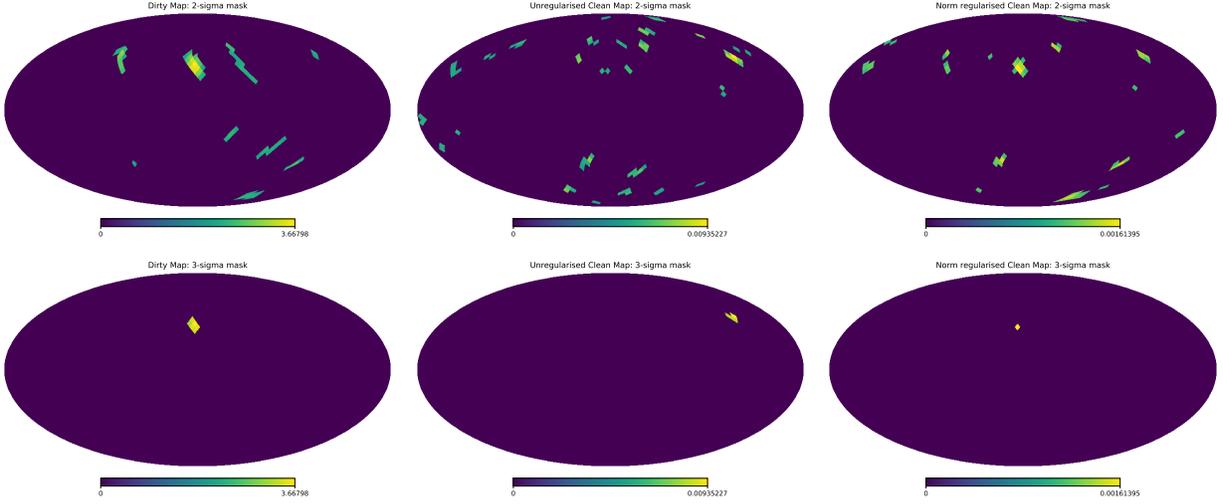

\centering
\includegraphics[width = 0.3\textwidth]{{{dirty-point0.05_2sigma-mask}}}
\includegraphics[width = 0.3\textwidth]{{{clean_no-reg_point0.05_iters3_2sigma-mask}}}
\includegraphics[width = 0.3\textwidth]{{{clean_norm-reg_point0.05_lambda1000_2sigma-mask}}}\\
\includegraphics[width = 0.3\textwidth]{{{dirty-point0.05_3sigma-mask}}}
\includegraphics[width = 0.3\textwidth]{{{clean_no-reg_point0.05_iters3_3sigma-mask}}}
\includegraphics[width = 0.3\textwidth]{{{clean_norm-reg_point0.05_lambda1000_3sigma-mask}}}
\caption{Plots of positive 2$\sigma$ (top row) and 3$\sigma$ (bottom row) outliers for dirty map (left) and clean maps with (right) and without (middle) regularisation, corresponding to the very weak point source injection shown in Fig.~\ref{fig:point}. Spurious localised sources appear in all the maps with a 2$\sigma$ mask. Even with a 3$\sigma$ mask dirty map and unregularalised clean map show spurious outliers, while the norm regularised clean map precisely locates the source pixel.}
\label{fig:maskedMaps}
\end{figure*}

To further demonstrate the benefits of regularization we present the result of masking the maps at $2\sigma$ and $3\sigma$ cutoff levels where $\sigma$ is the standard deviation of the respective maps. Masking is a procedure where pixel values of the map are set to zero if they are smaller than the cutoff value. The masking was performed for the clean map corresponding to the third column of Fig.~\ref{fig:point}. We see that a $3\sigma$ mask correctly localizes the injected source in the regularised map. The masked unregularised clean map picks up a wrong pixel as shown in Fig.~\ref{fig:maskedMaps}.  This exercise is useful for detecting `outliers'---significantly bright pixels in the skymap. If the deconvolution itself creates outliers, the usefulness of the exercise will be severely compromised. These results also signify that, although we are using NMSE as one of the possible quality measures for reconstruction (which is certainly better than NSP for point sources), it may not capture the full extent of advantage of regularisation.
%MASKING 2 sigma and 3 sigma
\subsection{Robustness of regularised deconvolution}
\label{sec:robust}
So far we have considered specific injections in this study. We now show how any reasonable choice of the regularisation constant and number of iterations generically improves the quality of deconvolution.

\subsubsection{Insensitive to the number of iterations}

\begin{figure*}
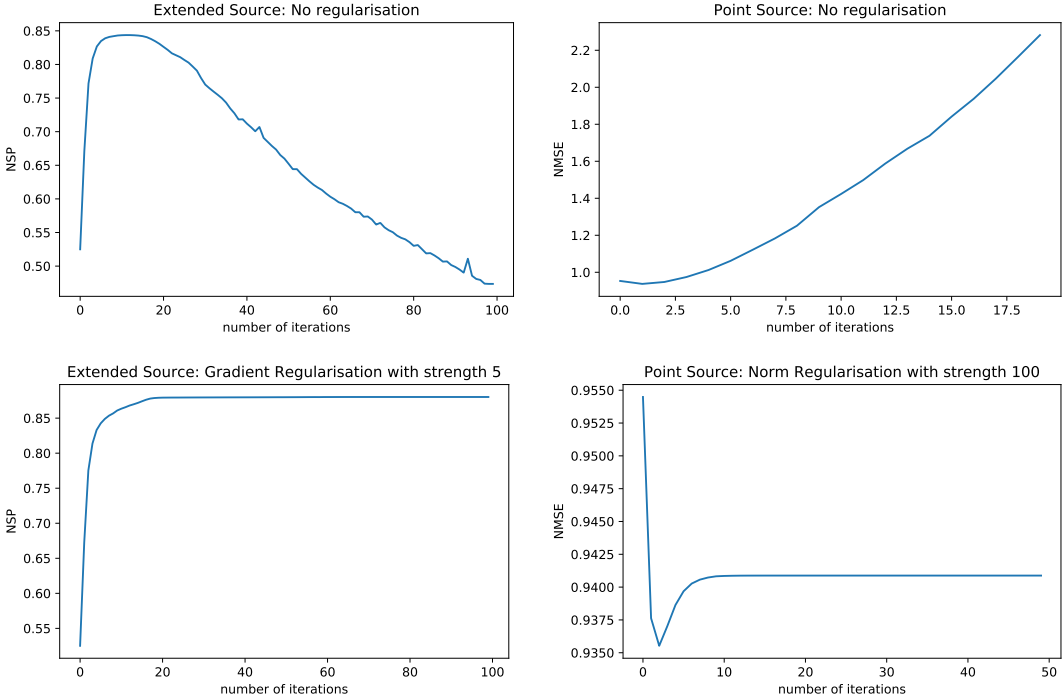

\centering
\includegraphics[width=0.4\textwidth]{{{nsp_no-reg_vs_iter_ext0.1}}}
\includegraphics[width=0.4\textwidth]{{{nmse_no-reg_vs_iter_point0.5}}}\\
\includegraphics[width=0.4\textwidth]{{{nsp_grad_vs_iter_ext0.1_lambda5}}}
\includegraphics[width=0.4\textwidth]{{{nmse_norm_vs_iter_point0.5_lambda100}}}
\caption{Stability of regularization with iteration number. The top two plots show how increasing the number of iterations deteriorates the quality of deconvolution for strong extended (left) and point (right) sources [these injections correspond to the left columns of Figures~\ref{fig:extended} and \ref{fig:point} respectively]. Regularisation stabilises the quality of deconvolution after $10-20$ iterations, as shown in the bottom plots for the corresponding sources with gradient and norm regularisations respectively. }
\label{fig:iter}
\end{figure*}

A considerable advantage of regularisation comes in the form of numerical stability of deconvolution. Quality of reconstruction of the SGWB map using unregularised deconvolution deteriorates after a few iterations due to the accumulation of numerical noise. This can be seen in the top panels of Fig.~\ref{fig:iter}. In the figure, NSP and NMSE are plotted against the number of iterations for a strong extended source (left) and a strong point-source (right) [same injections were used in the left column of Fig.~\ref{fig:extended} and \ref{fig:point} respectively]. The plots for regularised deconvolution are provided in the lower panels of Fig.~\ref{fig:iter}, which show that the quality of deconvolution stabilises after $\sim 10-20$ iterations for these cases. In accordance with this result, we choose to use $50-100$ iterations for regularised deconvolution to be on the safer side (though a smaller number could have yielded similar results).

\subsubsection{Nearly insensitive to the choice of $\lambda$}

Although we pick a value for the regularisation constant $\lambda$ for each injection separately such that regularisation produces optimal results, as seen in the Fig.~\ref{fig:nsp_gradient} and \ref{fig:nmse_norm}, we demonstrate that a broad range of values for $\lambda$ could produce similar results. Our results do not require a fine-tuned value of $\lambda$.  We find that any choice of $\lambda$ in a given range would produce fairly similar results as long as the strength of the source is similar, i.e., the procedure is not very sensitive to the exact shape and features of the source intensity distribution in the sky. For instance, $\lambda = 1000$ or $10000$ would produce acceptable results (within $\sim 10\%$ of the best reconstruction quality) for all the six injections considered in Fig.~\ref{fig:extended} and \ref{fig:point}. It is worth noting that stronger the regularisation strength, larger the bias. Also, strong gradient regularisation washes out the finer details, while a strong norm regularisation reduces the strength of the source. Although a visual inspection of our result confirms this aspect, it may not always be reflected in the NSP and NMSE measures. Therefore in order to capture the maximum information from the data, it is recommended that the smallest value of $\lambda$ that produces a reasonable reconstructed map be used. 

\subsubsection{Simulations}

We now demonstrate that the results are indeed insensitive to the choice of regularisation strength by performing simulations with a fixed choice of $\lambda$. We implement our regularized deconvolution routine on $1000$ simulations for each of the two cases - point sources and extended sources, corresponding to weak to very weak signal strength. For each simulation, we generate a dirty map by convolving the injected map with the beam and add a noise map. The noise map is generated by processing Gaussian noise in frequency domain corresponding to the two LIGO detectors. We follow the procedure described in \citet{Mitra07}.

\begin{figure*}
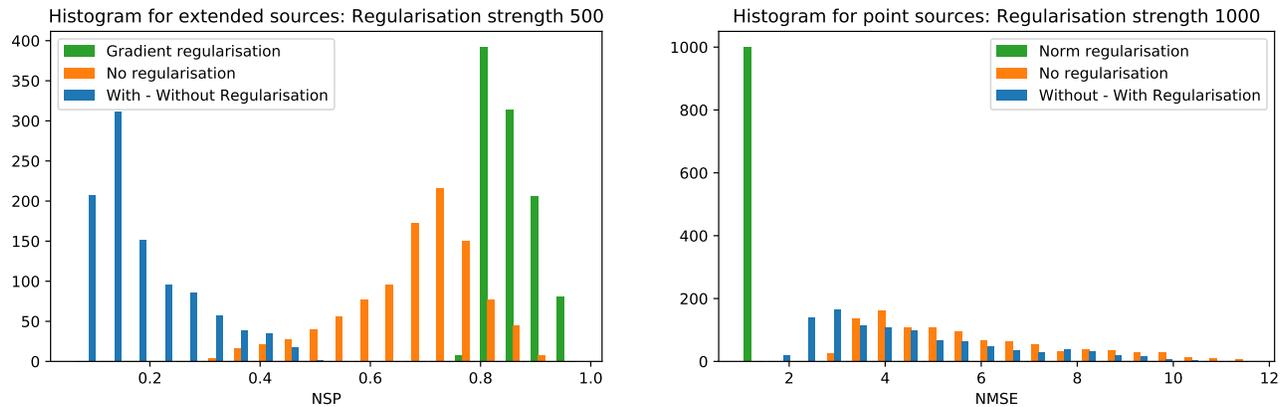

    \centering
    \includegraphics[width=\columnwidth]{{{hist_ext_grad_lambda500}}}
    \includegraphics[width=\columnwidth]{{{hist_point_norm_lambda1000}}}
    \caption{The above histograms show that regularized deconvolution gives better reconstruction compared to unregularized deconvolution even when the regularization constant is not fine-tuned for a specific strength or pattern. A simulation was performed with randomly chosen skymaps (left) or point sources injected in a random direction (right) with randomly chosen strengths of the injection in both the cases. Gradient regularisation was performed for extended sources and norm regularisation for point sources with fixed values of regularisation constant, $\lambda = 500$ and $1000$ respectively. Regularised clean maps (slanted hatched bars) have a significantly better quality of deconvolution compared to unregularised maps (horizontally hatched bars). In fact, the differences (unhatched bars) between NSP \& NMSE of regularised \& unregularised maps are always positive, implying that regularisation improved the quality of deconvolution in each of the simulations, not just statistically!}
    \label{fig:hist}
\end{figure*}

For point source injections, we randomly select a pixel and assign an intensity drawn from a uniform distribution in the `weak' to `very weak' range. We deconvolve the dirty map without regularisation and with norm regularisation with $\lambda = 1000$. We then make histograms of NMSE and its difference obtained from these two types of clean maps, as presented in the right panel of Fig.~\ref{fig:hist}.

The process is more elaborate for extended sources. We compute the angular power spectrum, $C_{l}$, of the Milky-Way-Galaxy-like map that we have been using in this paper. We then generate simulated maps from this $C_{l}$ using HEALPix tools. We take the absolute value of the maps to make all the pixels positive and finally multiply with random scaling factors uniformly drawn from a suitable range; the range is chosen such that the strength of the injections lies in the range `weak' to `very weak'. We then create the dirty maps and deconvolve them without regularisation and with gradient regularisation with $\lambda = 500$. We make histograms of NSP for these two types of clean maps and its difference, as shown in the left panel of Fig.~\ref{fig:hist}.

The histograms in Fig.~\ref{fig:hist} clearly show that regularised deconvolution (slanted hatched bars) significantly outperforms unregularised deconvolution (horizontally hatched bars). The histograms of differences (plain bars) of NSP or NMSE between with and without regularisation (sign chosen appropriately) do not show any negative values. This further unveils that regularisation improves reconstruction not only statistically, but for each individual simulation.
\subsection{Prescription for Real Data}

We have now demonstrated that regularisation improves the quality of deconvolution and it is robust against the choice of parameters, provided we have some idea about the strength of the signal we are trying to probe. Since no SGWB has been detected so far, we expect that very weak sources will be the primary targets of a search in the coming years. On the contrary, we were unable to faithfully reconstruct skymaps that are much weaker than the ones considered here [third column of Fig.~\ref{fig:extended} and \ref{fig:point}] irrespective of the regularisation method and parameters. Given this limitation, a practical approach would be to target the weakest case that fall in the realm of validity of our procedure. From this perspective, to apply the procedure on the data from the upcoming observing runs of the current detectors, we recommend finding the optimal strength of regularisation for the weakest sources through a set of simulated injections as prescribed in this work. If the normalisation of data, beams, and noise properties are similar to those used in our simulation, we expect the optimal strength to lie in the range $\lambda \sim 500-1000$.

One can in principle vary $\lambda$ over a large range of values and see if anything interesting stands out in the map. However, this may not be the best practice to follow for all-sky all-frequency searches~\cite{eric_ain, Ain:2018zvo}, where the number of maps is already very high (equal to the number of frequency bins), making it computationally challenging and significantly boosting the risk of discovering `look elsewhere' effects.

One could also argue if the dirty maps could be directly used without performing deconvolution at all. In fact, for certain cases in Table~\ref{tab:nspnmse}, values of NSP or NMSE are indeed better for dirty maps. However, there are difficulties which prevent us from taking such advantages. First, the quality of dirty maps is worse than the regularised maps for very weak sources, which are going to be our primary targets. Reading from Table~\ref{tab:nspnmse}, for the very weak extended source, $\text{NSP}_\text{dirty} = 0.67 < \text{NSP}_\text{grad} = 0.79$ and for the very weak point source, $\text{NMSE}_\text{dirty} = 14.47 > \text{NMSE}_\text{norm} = 1.29$. Second, one needs to have a source sky model to use the dirty maps directly. Regularised deconvolution {\em does not} need a specific source model, it uses generic features of the source. In order to look for a point source in the dirty map whose location is not known, perhaps the most reasonable way would be to find outliers in the map, like in Fig.~\ref{fig:maskedMaps}, where the regularised clean map outperforms the dirty map. Similarly, to find whether the dirty map in Fig.~\ref{fig:extended} has an embedded source map, assuming that we had a model for the map (the Milky-Way-like pattern in this case, perhaps from electromagnetic surveys), we could find the NSP and test its statistical significance (similar to `Matched Filtering' the sky). Even in such cases, one would have to worry about the accuracy and completeness of the model. In the absence of a specific source model, there is no obvious way to check if the dirty map embeds an extended source and infer its shape, and therefore one has to work with the clean map.

\section{Conclusion}
 \label{sec:concl}

Mapping an anisotropic stochastic gravitational wave background using data from ground-based detectors is becoming progressively important as detectors are breaking sensitivity barriers and new cosmological results are being published. The task however is challenging. One fundamental hurdle is that the matrix that connects the source sky to the data is somewhat ill-conditioned, making it non-trivial to deconvolve the filtered cross-spectral data from pairs of detectors, a.k.a. the dirty map. In this work, we demonstrated that regularized deconvolution provides a robust yet straightforward way to address this issue and the method can be readily applied to the current LIGO-Virgo analyses.

Motivated by an earlier work on gravitational lensing~\cite{suyu}, we introduced and applied regularised deconvolution in SGWB mapmaking. We use two forms of regularisation function here: (1) norm regularisation that tries to minimise power in the whole map, which is suitable for localised sources (2) gradient regularisation that tries to reject small angular scale variations, which is suitable for extended sources. We use a Bayesian analysis to determine the optimal strength of regularization for the above two functions. The merits of these regularisation schemes are demonstrated using multiple examples of different source distributions in this paper. We show that regularisation dramatically improves the quality of reconstruction for weak sources, which are likely to be the primary candidates for the first detection. The method is not sensitive to a specific choice of the strength of regularisation, as long as the strength is chosen from a broad range of values determined by following the prescription given here. For strong sources regularisation is ineffective and sometimes worsens the quality of reconstruction compared to unregularised deconvolution. However, irrespective of the strength of the signal, regularisation stabilises the quality of deconvolution against iterations, making it a safer practice to follow.

In the detection problem, one is often interested to know the presence or absence of a particular known pattern in the data. In such a case, the likelihood ratio is the optimal detection statistics \cite{Talukder:2010yd}, where it is assumed that the pattern in the sky is accurately known. However, most often in the real scenarios, only a piece of information about the source intensity distribution is known, and it becomes vital to be able to regulate the strength of the prior. Implementing regularization in a Bayesian framework provides this versatility.

Our method is best suited for blind searches where the source location or intensity distribution is unknown, but we can use non-informative priors like gradient and norm regularization. It is worth mentioning here that unlike the detection problem where one is interested to know if a particular pattern is present in the data, the mapmaking problem concerns with reconstruction of the most probable pattern in the sky given the data. One can, in principle, use this reconstructed map to test the presence or absence of a specific pattern, but it may be more optimal to tailor an informative prior in such cases. The clean maps generated by our method are readily useful to search for anomalies and outliers caused by {\em unknown} sources or instrumental artifacts. It is particularly relevant for conducting blind all-sky narrowband searches~\cite{eric_ain}, though we have not yet tested our method for that application, which may pose a computational challenge if not implemented carefully.

Here we have considered only the LIGO Hanford and Livingston detectors. The method can be easily extended to a network of detectors by replacing the beam matrix and dirty map by their sums from the individual baselines (with carefully chosen normalisation)~\cite{Mitra07,Talukder:2010yd,PhysRevD.80.122002}. The network acts as a natural regulariser to a certain extent~\cite{Talukder:2010yd,PhysRevD.80.122002}, so it may require a smaller regularisation strength for the same pixel resolution. However, when KAGRA and LIGO-India join the network, the baseline lengths will significantly increase, leading to a much higher resolving power. This will demand finer pixels (increasing the number of pixels by a factor of 4 or more), making it challenging to invert the beam matrix, possibly necessitating an even stronger regularisation.

The stochastic searches are routinely being conducted in Spherical Harmonic basis~\cite{LIGOScientific:2019gaw}; deconvolution is a big challenge in this basis as well~\cite{PhysRevD.80.122002}. While for the case of pixel-based radiometer analysis no regularised deconvolution has been conventionally used, for the case of spherical harmonic analysis a heuristic SVD-based regularisation has been implemented. Based on the advantage we gain by using Bayesian regularization in pixel basis, we are encouraged to propose a similar approach to the spherical harmonic basis as well. 
                                                        
In the recent past, many theoretical models have been introduced to predict anisotropic SGWB from compact binaries ~\cite{Jenkins:2018uac,Jenkins:2018uac,Cusin:2018rsq,Cusin:2017mjm,Cusin:2017fwz} and Nambu-Goto Cosmic Strings~\cite{Jenkins:2018lvb}. These models predict an angular power spectrum of the sky, $C_l$, with part of the power accessible in the sensitive frequency band of the ground-based laser interferometric detectors, which will allow constraining of these models using data from those detectors through the estimation of $C_l$~\cite{LIGOScientific:2019gaw}. Since the dirty maps are convolution of the source with an extended asymmetric beam, estimation of $C_l$ of the true sky, will require a reliable clean map where our method should prove to be handy. The improvement of NSP through regularisation for extended sources should help studies performing cross-correlation between SGWB maps and the large scale structures. Going beyond, since the regularization scheme applied in this paper for SGWB mapmaking is fairly generic, we hope that it will open new avenues to aid the above mentioned and other related investigations in the future.

\begin{acknowledgments}
We would like to thank Sherry H. Suyu and Vuk Mandic for suggesting and discussing this possibility many years ago! We also thank Joe Romano for his valuable comments. This work significantly benefitted from the interactions with the Stochastic Working Group of the LIGO-Virgo Scientific Collaboration. We acknowledge the use of IUCAA LDAS cluster Sarathi for the computational/numerical work. This research benefited from a grant awarded to IUCAA by the Navajbai Ratan Tata Trust (NRTT). SM acknowledges support from the Department of Science and Technology (DST), India provided under the Swarna Jayanti Fellowships scheme. JS acknowledges the support by JSPS KAKENHI Grant Number JP17H06361. This article has a LIGO document number LIGO-P1800278 and IUCAA Preprint number IUCAA-07/2019. SB acknowledges the financial support provided by the SUGWG group at the Syracuse University under the grant number PHY-1707954 and the DARKGRA group in La Sapienza under the European Union's H2020 ERC, Starting Grant agreement no.~DarkGRA--757480 and  support from the Amaldi
Research Center funded by the MIUR program ``Dipartimento di Eccellenza'' (CUP: B81I18001170001).
\end{acknowledgments}

\appendix

\section{Derivation of expression for $\mathcal{S}_{mp}$}
\label{app:smp}

The most probable solution is defined as, 
\begin{align}
    \nabla_{S} M(\mathcal{S}_{mp})= \nabla_{S} \chi^{2}(\mathcal{S}_{mp}) + \mu \nabla_{S} R(\mathcal{S}_{mp}) = 0.  
    \label{eq:M-defined}
\end{align}
In this section we prove that the $\mathcal{S}_{mp}$ can be calculated as $\mathcal{S}_{mp} = (\mathcal{B} + \lambda \mathcal{C})^{-1} \mathcal{D}$.

Let $R(\mathcal{S})$ be any quadratic function of $\mathcal{S}$ with a minima located at $\mathcal{S}_{reg}$ and let $\mathcal{C}$ be the hessian. By Taylor expanding about the minima $\mathcal{S}_{reg}$,  $R(\mathcal{S})$ can be written as, 
\begin{align}
    R(\mathcal{S})= \mathcal{S}_{reg} + \frac{1}{2} \mathcal{S}^{T} \mathcal{C} \mathcal{S}
\end{align}
For both the regularization functions we consider in our study, $\mathcal{S}_{reg} = 0$, and, therefore, 
\begin{align}
    R(\mathcal{S})=  \frac{1}{2} \mathcal{S}^{T} \mathcal{C} \mathcal{S} \, .
\label{eq:Rs-defined}
\end{align}
Now, recall that $\mathcal{N} = \kappa \mathcal{B}$ and $\lambda = \mu \kappa$. Then, 
\begin{align}
\begin{split}
    \chi^{2} \ \equiv\ & \frac{1}{2} \, (\mathcal{B S - D})^{T} \mathcal{N}^{-1} (\mathcal{B S - D)}  \\ 
    =\ & \frac{\mu}{2\lambda} \, ( \mathcal{B S -D})^{T} \mathcal{B}^{-1} (\mathcal{B S - D}) \, .
\end{split}
\label{eq:def-chi2}
\end{align}
Next, we compute the functional form of $\nabla_{S} M(\mathcal{S})$. 
\begin{align}
    \begin{split}
       & \nabla_{S} M(\mathcal{S}) \ =\ \nabla_{S} \chi^{2} (\mathcal{S}) + \mu \nabla_{S} R( \mathcal{S}) \\
       & \ =\ \frac{\mu}{2\lambda} \, \nabla_{S} \left[ ( \mathcal{B S -D})^{T} \mathcal{B}^{-1} (\mathcal{B S - D}) \right] + \frac{\mu}{2}\nabla_{S} \left[\mathcal{S}^{T} \mathcal{C} \mathcal{S}\right] \\
       & \ =\  (\mu/\lambda) (\mathcal{B S - D}) + \mu ( \mathcal{C S} ) \, .
    \end{split}
\label{eq:setting-nabla}
\end{align}
To solve for $\mathcal{S}_{mp}$, we set $\nabla_{S} M(\mathcal{S}_{mp}) = 0$ and get,
\begin{equation}
    \mathcal{B} \mathcal{S}_{mp} \ + \ \lambda \mathcal{C} \, \mathcal{S}_{mp} \ = \ \mathcal{D} \, ,
\end{equation}
that is,
\begin{equation}
    \mathcal{S}_{mp} \ = \ (\mathcal{B} \, + \, \lambda \mathcal{C})^{-1} \mathcal{D} \,, \label{eq:smp_app}
\end{equation}
the most probable solution.

\section{Bias Introduced by Regularization}
\label{app:bias}

The map reconstructed using regularization is a biased estimator while the un-regularized reconstruction is unbiased. 
Bias $\wp$ is defined, 
\begin{equation}
\label{eq:bias}
    \wp = \mathcal{S}_{\mathrm{true}} - <\hat{\mathcal{S}}> \, .
\end{equation} 
For unregularized reconstruction,
\begin{equation}
<\hat{\mathcal{S}}> = ( \mathcal{B}^T \mathcal{N}^{-1} \mathcal{B})^{-1} ( \mathcal{B}^T \mathcal{N}^{-1}\mathcal{B}) \mathcal{S}_{true} = \mathcal{S}_{true} \,.
\end{equation}
From the above equation, it is clear that $\wp = 0$ for the unregularized clean map. Regularization introduce the bias in the reconstructed map by altering the pixel-to-pixel covariance matrix of the reconstructed map.

For the case of regularized deconvolution, from Eq.~(\ref{eq:smp_app}) the bias $\wp$ can be calculated as,
\begin{equation}
    <\mathcal{S}_{mp}> \ = \ (\mathcal{B} + \lambda \mathcal{C})^{-1} <\mathcal{B} \cdot \mathcal{S} + \mathbf{n}> = (\mathcal{B}+\lambda \mathcal{C})^{-1} \mathcal{B} \cdot \mathcal{S}_{\mathrm{true}} \,,
\end{equation}
Hence, the bias $\wp$ for this case is,
\begin{equation}
\wp = \mathcal{S}_{\mathrm{true}} -<\hat{\mathcal{S}}> = (\mathbf{1}-(\mathcal{B}+\lambda \mathcal{C})^{-1} \mathcal{B}) \cdot \mathcal{S}_{\mathrm{true}} \, .
\end{equation}

In the case where the source intensity is strong, regularization unnecessarily introduces a bias, and therefore, an unregularized deconvolution performs slightly better. The bias cannot be corrected, because it requires the knowledge of the true sky which is being estimated.

\bibliography{regDeconvSGWB.bib}

\end{document}